\documentclass[showpacs,aps,twocolumn]{revtex4}
\usepackage{epsfig}
\usepackage{graphicx}
\usepackage{amsmath,amssymb,amsfonts}
\usepackage{amsmath}
\usepackage{array}
\usepackage{url}
\usepackage{multirow}
\usepackage{float}
\usepackage{comment}

\usepackage{booktabs}  % for \toprule, \bottomrule
\usepackage{longtable} % if the table is too long for one page

\usepackage{lineno}
\usepackage{xspace}
\usepackage{ulem}

\newcommand{\bef}{\begin{figure}}
\newcommand{\eef}{\end{figure}}
\newcommand{\bc}{\begin{center}}
\newcommand{\ec}{\end{center}}

\newcommand{\be}{\begin{equation}}
\newcommand{\ee}{\end{equation}}
\newcommand{\bea}{\begin{eqnarray}}
\newcommand{\eea}{\end{eqnarray}}

\def\ba{\begin{eqnarray}}
\def\ea{\end{eqnarray}}

\usepackage[usenames,dvipsnames]{color}
\definecolor{darkblue}{RGB}{0,0,196}
\usepackage[colorlinks=true,linkcolor=darkblue,citecolor=darkblue,urlcolor=darkblue]{hyperref}

\begin{document}
\title{Barnett effect as a new source of magnetic field in heavy-ion collisions}

\author{Dushmanta Sahu}
\email{Dushmanta.Sahu@cern.ch}
\affiliation{Instituto de Ciencias Nucleares, Universidad Nacional Autónoma de México, Apartado Postal 70-543,
México Distrito Federal 04510, México}

\begin{abstract}

The Barnett effect is a fundamental magnetomechanical phenomenon in which a ferromagnetic material becomes magnetized under rotation. Using a hadron resonance gas (HRG) model under rigid rotation, we compute the Barnett magnetization ($M_{\rm Barnett}$) and show that it produces a magnetic field ($B_{\text{ind}}$) comparable in magnitude to the well-known external field ($B_{\text{ext}}$) from spectator protons at low energy heavy-ion collisions. This finding establishes the Barnett effect as a previously overlooked but essential source of magnetization and magnetic field in the heavy-ion collisions, with profound implications for understanding spin dynamics and anomalous transport in quantum chromodynamics under extreme rotation.

\end{abstract}

\maketitle

\section{Introduction}
\label{sec1}

In 1915, Samuel Barnett discovered a unique phenomenon where a ferromagnetic material becomes magnetized under rotation~\cite{Barnett:1915uqc}. Later termed the Barnett effect, this magnetomechanical coupling is fundamental to condensed matter physics and spintronics. It is caused by the coupling between the angular momentum of electronic spins and the rotational motion of the ferromagnetic rod. Recently, Arabgol and Sleator observed the nuclear Barnett effect for the first time by rotating a water sample at speeds up to $13.5$ kHz in a weak magnetic field. They measured a change in proton polarization proportional to the rotation frequency~\cite{NBE}. This motivates the intriguing possibility that similar magnetomechanical effects could manifest in relativistic many-body systems, such as quantum chromodynamics (QCD) matter created in heavy-ion collisions. Non-central heavy-ion collisions at the Relativistic Heavy Ion Collider (RHIC) and the Large Hadron Collider (LHC), create unique conditions. In such environments, fast-moving spectator protons generate transient, ultra-strong magnetic fields ($eB \sim 10^{15}$--$10^{19}$ G)~\cite{Kharzeev:2007jp}, while the large angular momentum of the colliding nuclei produces rapid global rotation ($\omega \sim 10^{21}$ rad/s)~\cite{Becattini:2007sr}. The combination of these strong magnetic fields and vorticity creates a novel environment for studying spin dynamics in QCD matter. While most attention has focused on ultra-high energies (top RHIC and LHC), the low-energy regime explored by the RHIC Beam Energy Scan (BES) program presents a unique opportunity. Here, the baryon density is high, enhancing matter-antimatter asymmetry, while the initial magnetic field $B_{\text{ext}}$ is significantly weaker. In this specific regime, we predict that the rotation-induced Barnett effect can become a significant mechanism for spin polarization, also fundamentally altering the interpretation of other experimental signals.

Researchers have recently begun extensively studying the impact of external magnetic fields and rotation on hot QCD matter, as both significantly modify its thermodynamics~\cite{Endrodi:2013cs,Kadam:2019rzo,Pradhan:2021vtp,Sahoo:2023vkw} and transport properties~\cite{Dash:2020vxk,Dey:2020awu}. Lattice QCD (lQCD) has explored QCD matter under magnetic fields in detail~\cite{Endrodi:2015oba,Endrodi:2024cqn}, and under rotation to a lesser extent~\cite{Yamamoto:2013zwa,Braguta:2023iyx}. These conditions also induce anomalous phenomena like chiral magnetic effect (CME)~\cite{Fukushima:2008xe} and chiral vortical effect (CVE)~\cite{Shitade:2020lfe}. Both stem from the chiral anomaly in the quark-gluon plasma (QGP)~\cite{Kharzeev:2010gr} and are probed experimentally via charge-dependent azimuthal correlations and flow. Extreme rotation also polarizes hadrons, particularly $\Lambda$ hyperons, via spin–vorticity coupling~\cite{STAR:2023nvo,STAR:2017ckg}. Experiments test all these effects through CME-sensitive correlations~\cite{ALICE:2022ljz}, $\Lambda$ polarization, and flow observables. Ongoing RHIC and LHC measurements continue to advance our understanding of QCD under such extreme conditions. In addition to the above-mentioned effects, the medium formed in these collisions may exhibit a remarkable analog of the Barnett effect. While the effects of external magnetic fields and vorticity have been studied separately, the magnetization arising from vorticity itself, the Barnett effect, and its feedback on system dynamics has not been systematically investigated in the QCD context. In this scenario, particles with spin become magnetized along the rotation axis due to the coupling between their intrinsic angular momentum and the vorticity field. The Barnett mechanism in this context differs fundamentally from polarization induced by fluid vorticity or magnetic fields alone. It depends critically on the magnetic moments ($\mu_{m}$) of hadrons, leading to species-dependent alignment patterns. 

To investigate this potential Barnett magnetization ($M_{\rm Barnett}$) in the hadronic phase, we employ the hadron resonance gas (HRG) model under rotation. The HRG model provides a successful framework for describing the hadronic phase, reproducing the hadron yields~\cite{Andronic:2005yp} and matching lattice QCD thermodynamics results for temperatures up to $T \sim 150$ MeV at vanishing baryochemical potential~\cite{Karsch:2003vd}. Its versatility extends to systems under extreme conditions, including external magnetic fields~\cite{Kadam:2019rzo,Pradhan:2021vtp}, with results agreeing well with lQCD. While lQCD breaks down at high $\mu_{B}$ due to the fermion sign problem~\cite{HotQCD:2014kol}, the HRG model provides robust estimations of thermodynamic quantities even at extreme densities. Recently, researchers have extended the HRG model to study the effect of rotation in the medium~\cite{Fujimoto:2021xix}. Such studies predict a deconfined phase transition at high angular velocity. Subsequently, the rotating HRG model has also been used to study the thermodynamic and transport properties of hadronic matter~\cite{Pradhan:2023rvf,Mukherjee:2023ijv,Sahoo:2025fif,Dwibedi:2024amt,Padhan:2024edf}. We emphasize that the use of the ideal HRG model is a deliberate choice to provide a baseline, first-principles calculation of the pure Barnett contribution, isolating it from other complex hadronic interactions. While interactions (e.g., excluded volume effects) will modify the absolute magnitude of the pressure and thus the magnetization, the fundamental qualitative trend, monotonic increase with $T$, $\mu_B$, and $\omega$, and the order-of-magnitude estimate for $B_{\text{ind}}$ are expected to be robust. This approach allows us to establish a clear, conservative estimate of rotation-induced magnetization, which can later be refined with more sophisticated interaction models. By employing the rotating HRG model, we quantify Barnett magnetization and highlight its capacity to generate a new previously overlooked magnetic field, $B_{\rm ind}$, which can potentially be of the same order as the external magnetic field produced in low energy heavy-ion collisions.

\section{Formulation}
\label{sec2}

The ideal HRG model assumes that the hadronic matter is in thermal equilibrium, and all the hadronic states are point-like particles with no interactions between them. The grand canonical partition function in the HRG model is defined as~\cite{Andronic:2012ut},

\begin{equation}
{\rm ln}Z_{i}^{id} = \pm\frac{Vg_{i}}{2\pi^{2}} \int_{0}^{\infty} p_{i}^{2}dp_{i} ~{\rm ln}(1\pm exp[-(E_{i} - \mu_{i})/T])
\end{equation}
 where, $g_{i}$, $p_{i}$, $m_{i}$ and $E_{i} = \sqrt{p_{i}^{2} + m_{i}^{2}}$ are the degeneracy, momentum, mass and energy of the $i$th hadron respectively. The $\pm$ signs correspond to fermions and bosons in the system, and $\mu_{i}$ is the total chemical potential of the system given as,
\begin{equation}
    \mu_{i} = B\mu_{\rm{B}} + S\mu_{\rm{S}} + Q\mu_{\rm{Q}},
\end{equation}
where $B$, $S$, and $Q$ are baryon, strangeness, and electric charge quantum numbers, respectively. The pressure $P_{i}$ in HRG is given as,

 \begin{equation}
     P_{i} = \pm \frac{Tg_{i}}{2\pi^{2}} \int_{0}^{\infty} p_{i}^{2}dp_{i} ~{\rm ln}(1\pm exp[-(E_{i} - \mu_{i})/T])
 \end{equation}

For a HRG system under rotation, the pressure for ith hadron under rotation is given by~\cite{Fujimoto:2021xix},

\begin{multline}
P_{i} = \pm \frac{T}{8\pi^2} \int_{(\xi_{l,1}\omega)^2} dp_r^2 \int dp_z \sum_{l=-\infty}^{\infty} \sum_{v=l}^{l+2S_i} 
\\
J_v^2(p_r r) 
\ln\left(1 \pm e^{-(E_{i} - (l + S_i)\omega - \mu_i)/T}\right)
\end{multline}

where, $E_{i} = \sqrt{p_r^2 + p_z^2 + m_i^2} - (l+S_{i})\omega$ is the energy spectrum relation, where $l$ is the orbital angular momentum, $S_{i}$ is the spin of ith hadron and $J_v(p_r r)$ is the Bessel function. $\xi_{l,1}$ is the first zero of Bessel function $J_{0}$.

\subsubsection{Rotation to Effective Field: Barnett Magnetization}

The Barnett effect arises from the equivalence between the energy of a magnetic moment in a magnetic field and the energy of a spin in a rotating frame. We derive the effective magnetic field \(B_{\text{eff}} \) by equating these two energy shifts. The interaction energy for a magnetic moment $\mu_{m}$ in an external magnetic field \( \mathbf{B} \) is,
\[
E_{\text{magnetic}} = - \mu_{m} \cdot \mathbf{B}
\]
The magnetic moment is related to the spin operator \( \mathbf{S} \) by,
\[
\mu_{m} = g \left( \frac{\mu_N}{\hbar} \right) \mathbf{S},
\]
where \( g \) is the dimensionless \(g\)-factor and \( \mu_N \) is the nuclear magneton. Assuming the field is along the \(z\)-axis, \( \mathbf{B} = (0, 0, B_z) \), the energy shift becomes,
\[
\Delta E_{\text{magnetic}} = - \mu_{m,z} B_z = - g \left( \frac{\mu_N}{\hbar} \right) S_z B_z.
\]
For a quantum state with a definite spin projection \( s_z \) (an eigenvalue of \( S_z \)), the energy shift is,
\begin{equation}
\Delta E_{\text{magnetic}} = - g \left( \frac{\mu_N}{\hbar} \right) s_z B_z.
\label{eq:Zeeman}
\end{equation}

In a frame rotating with angular velocity \( \bm{\omega} \) around the \(z\)-axis, the Hamiltonian acquires a term coupling to the total angular momentum. For an elementary particle, this is its spin,
\[
H_{\text{rotation}} = - \bm{\vec\omega} \cdot \mathbf{S}.
\]
Assuming \( \bm{\vec\omega} = (0, 0, \omega_z) \), the energy shift for a state with spin projection \( s_z \) is,
\begin{equation}
\Delta E_{\text{rotation}} = - \omega_z s_z.
\label{eq:Rotation}
\end{equation}

The fundamental premise of the Barnett effect is that the physical consequences of rotation are identical to those of a magnetic field. This requires the energy shifts for any spin state to be equal;
\[
\Delta E_{\text{rotation}} = \Delta E_{\text{magnetic}}.
\]
Substituting equations (\ref{eq:Rotation}) and (\ref{eq:Zeeman}) yields,
\[
- \omega_z s_z = - g \left( \frac{\mu_N}{\hbar} \right) s_z B_z.
\]
The spin projection \( s_z \) cancels out, giving
\[
\omega_z = g \left( \frac{\mu_N}{\hbar} \right) B_z.
\]

Solving for \( B_z \) gives the strength of the effective magnetic field that mimics the rotation,
\[
B_z = \frac{\hbar \omega_z}{g \mu_N}.
\]
We denote this effective field as \( B_{\text{eff}} \), leading to the final result in natural units as,
\begin{equation}
B_{\text{eff}} = \frac{\omega}{g \mu_N},
\end{equation}
where \( \omega \) is the magnitude of the angular velocity and the sign of \( g \) determines the direction of the field relative to the rotation axis.

 The orbital angular momentum is neglected because the equivalence is derived for the intrinsic spin of a stationary particle, for which the orbital angular momentum is zero. This is fundamental as the magnetic Hamiltonian couples only to the spin, and we compare it to the part of the rotational interaction that does the same. Orbital effects of rotation are physically separate and are incorporated into the system's pressure, $P(T, \mu, \omega)$.

Now, a hadron gas when subjected to finite rotation develops Barnett magnetization, $M_{{\rm Barnett},i}$ which responds to the effective magnetic field $B_{\text{eff}}$. Thus, the Barnett magnetization is given as,

\begin{equation}
M_{Barnett,i} = \left(\frac{\partial P}{\partial B_{\text{eff}}}\right)_{T,\mu} = g_i \mu_N\left(\frac{\partial P}{\partial \omega}\right)_{T,\mu}
\end{equation}

Here, $P = \sum_{i} P_{i}$ is the total pressure of the system. At $\mu=0$, matter and antimatter contributions cancel due to opposite magnetic moments, leading to vanishing net Barnett magnetization. Only with finite $\mu_B$ does a matter–antimatter asymmetry survive, allowing for finite magnetization.

\subsubsection{Barnett Magnetization to Induced Magnetic Field}

 The magnetization induced due to rotation will further generate a magnetic field, which we define as $B_{\rm ind}$, which will further influence the system. A key distinction to stress here is that while $B_{\rm eff}$ is a fictitious field which encodes the spin-rotation coupling, $B_{\rm ind}$ is a real and physical field that is a consequence of the Barnett magnetization. The latter is the field that can influence spin dynamics, transport phenomena, and can, in principle, be measured indirectly. The single-particle energy eigenvalue for a particle in a rotating frame, accounting for the Barnett-effect-induced magnetic field $B_{\text{ind}}$, is given by;
\[
E = \sqrt{p_z^2 + p_T^2 + m^2} - (l + s_z)\omega  - \mu_{m} \cdot B_{\text{ind}}
\]

where, $\mu_{m}$ is the magnetic moment value. For a uniformly magnetized object with no free currents and no demagnetizing field, the induced magnetic field can be given as~\cite{griffiths2017electrodynamics},
\[
B_{\text{ind}} = \mu_0 M,\]

 where, $\mu_{0}$ is the magnetic permeability in vacuum, taken as 1 in natural units. This relation holds true for an infinitely long cylinder magnetized along its axis. We employ it here as a well-justified approximation for the fireball created in a peripheral heavy-ion collision, where the global vorticity, and hence the Barnett magnetization, is aligned with a preferred direction (approximately the beam axis). Furthermore, this choice provides a direct and conservative estimate of the field strength generated by the magnetization itself. A more detailed calculation incorporating the exact, time-evolving geometry and demagnetization effects is not expected to alter our main qualitative conclusion regarding the significance of the Barnett-induced field.

Furthermore, as shown in Ref.\cite{Cleymans:2005xv}, we use the parametrization which allows us to study the magnetization as a function of center-of-mass energy. The parametrization which is well-established is given as, 
\begin{equation*}
    T(\mu_{B}) = q_{1} - q_{2}\mu_{B}^{2} - q_{3}\mu_{B}^{4},
\end{equation*}
\begin{equation*}
    \mu_{B}(\sqrt{s_{\text{NN}}}) = \frac{q_{4}}{1+q_{5}\sqrt{s_{\text{NN}}}},
\end{equation*}
where, $q_{1} = 0.166~\text{GeV}$, $q_{2} = 0.139~\text{GeV}^{-1}$, $q_{3} = 0.053~\text{GeV}^{-3}$, $q_{4} = 1.308~\text{GeV}$, and $q_{5} = 0.273~\text{GeV}^{-1}$. 
These parameters are obtained using freeze-out criteria based on the ideal HRG model.

\section{Results and Discussion}
\label{sec3}

\begin{figure*}[ht!]
    \includegraphics[width = 0.48\linewidth]{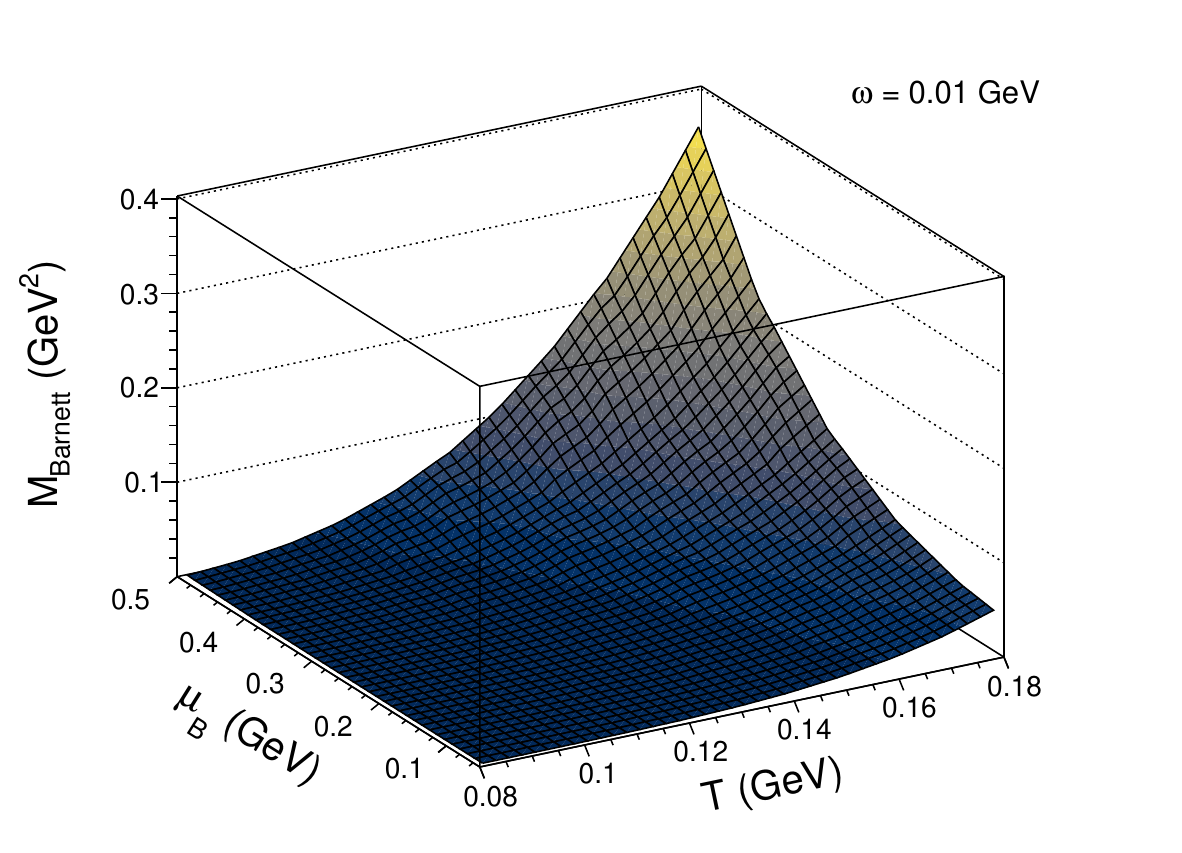}
        \includegraphics[width = 0.48\linewidth]{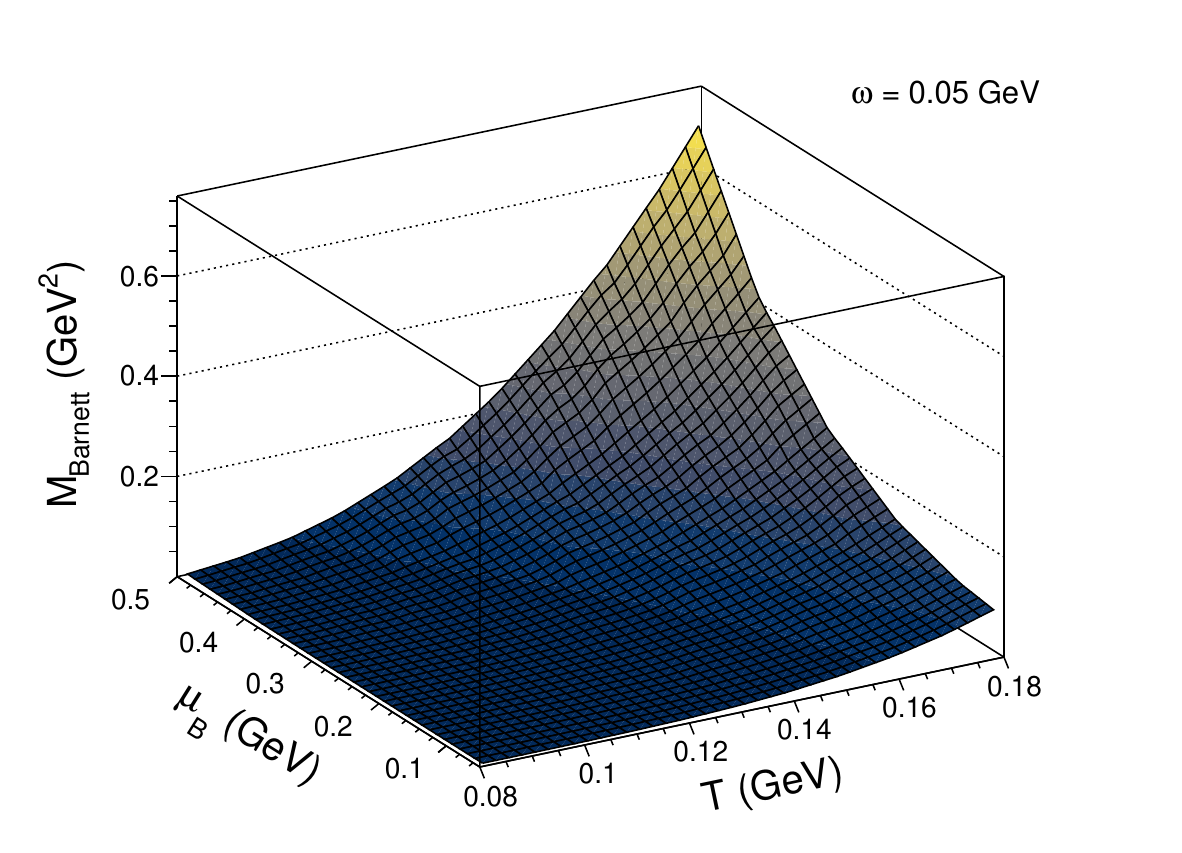}
    \caption{Barnett magnetization as a function of temperature and baryochemical potential for $\omega$ = 0.01 GeV (left panel) and 0.05 GeV (right panel).}
        \label{fig1}
\end{figure*}

\begin{comment}

\begin{figure*}[ht!]
    \includegraphics[width = 0.45\linewidth]{M_vs_T.pdf}
    \includegraphics[width = 0.45\linewidth]{M_vs_mu.pdf}
    \caption{Barnett magnetization as a function of temperature for different $\omega$ values at a constant baryochemical potential of 0.3 GeV (left panel) and Barnett magnetization as a function of baryochemical potential for different $\omega$ values at a constant temperature of 0.15 GeV (right panel).}
        \label{fig3}
\end{figure*}

\end{comment}

\begin{comment}

\begin{figure}[ht!]
    \includegraphics[width = 0.99\linewidth]{2D_M.pdf}
    \caption{Barnett magnetization as a function of center of mass energies for two different $\omega$ values.}
        \label{fig4}
\end{figure}
\end{comment}

\begin{figure}[ht!]
    \includegraphics[width = 0.90\linewidth]{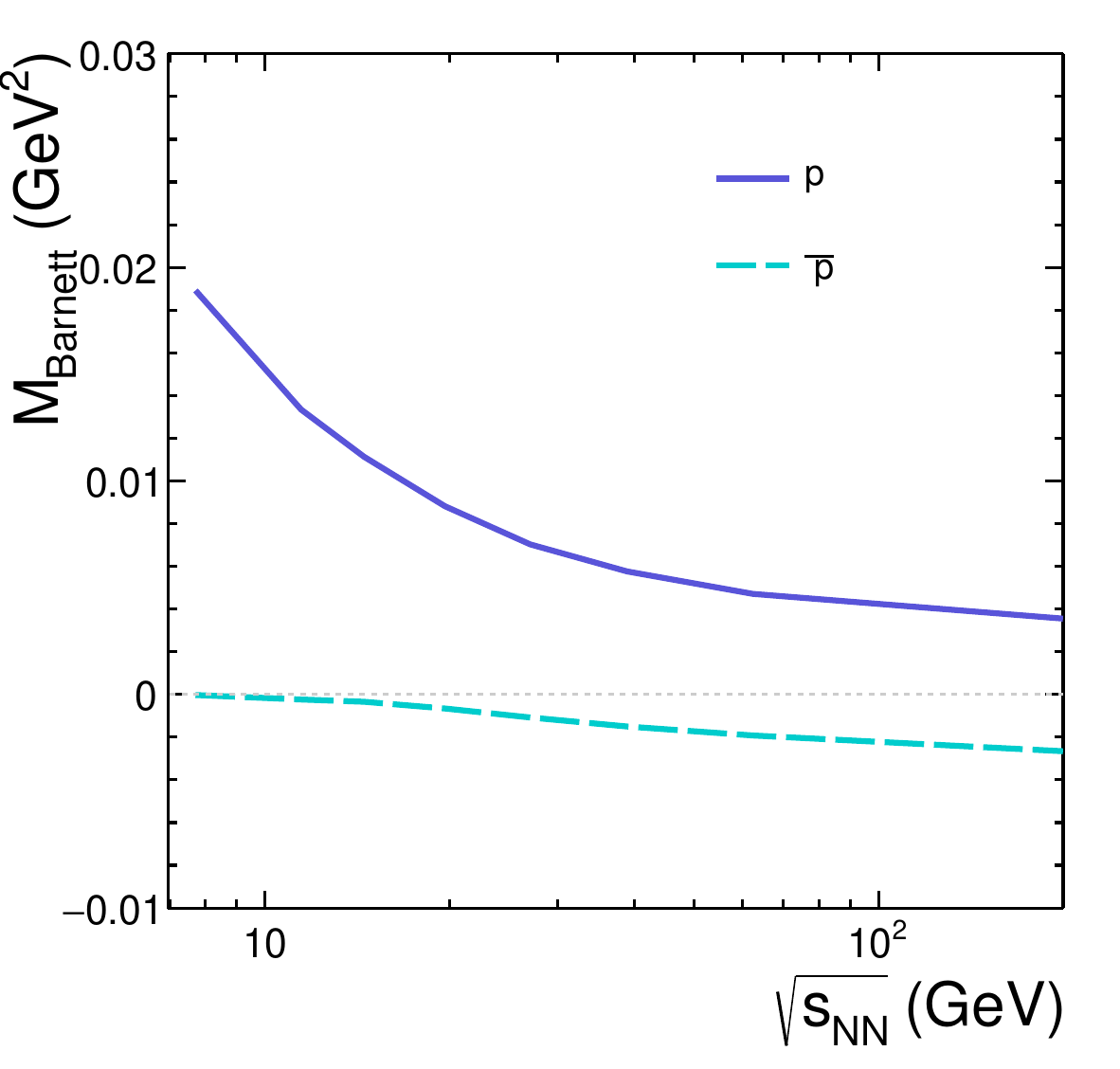}
    \caption{Barnett magnetization of proton and anti-proton as a function of center of mass energies for the $\omega$ values taken from Ref.~\cite{STAR:2017ckg}.}
        \label{fig2}
\end{figure}

\begin{figure}[ht!]
    \includegraphics[width = 0.90\linewidth]{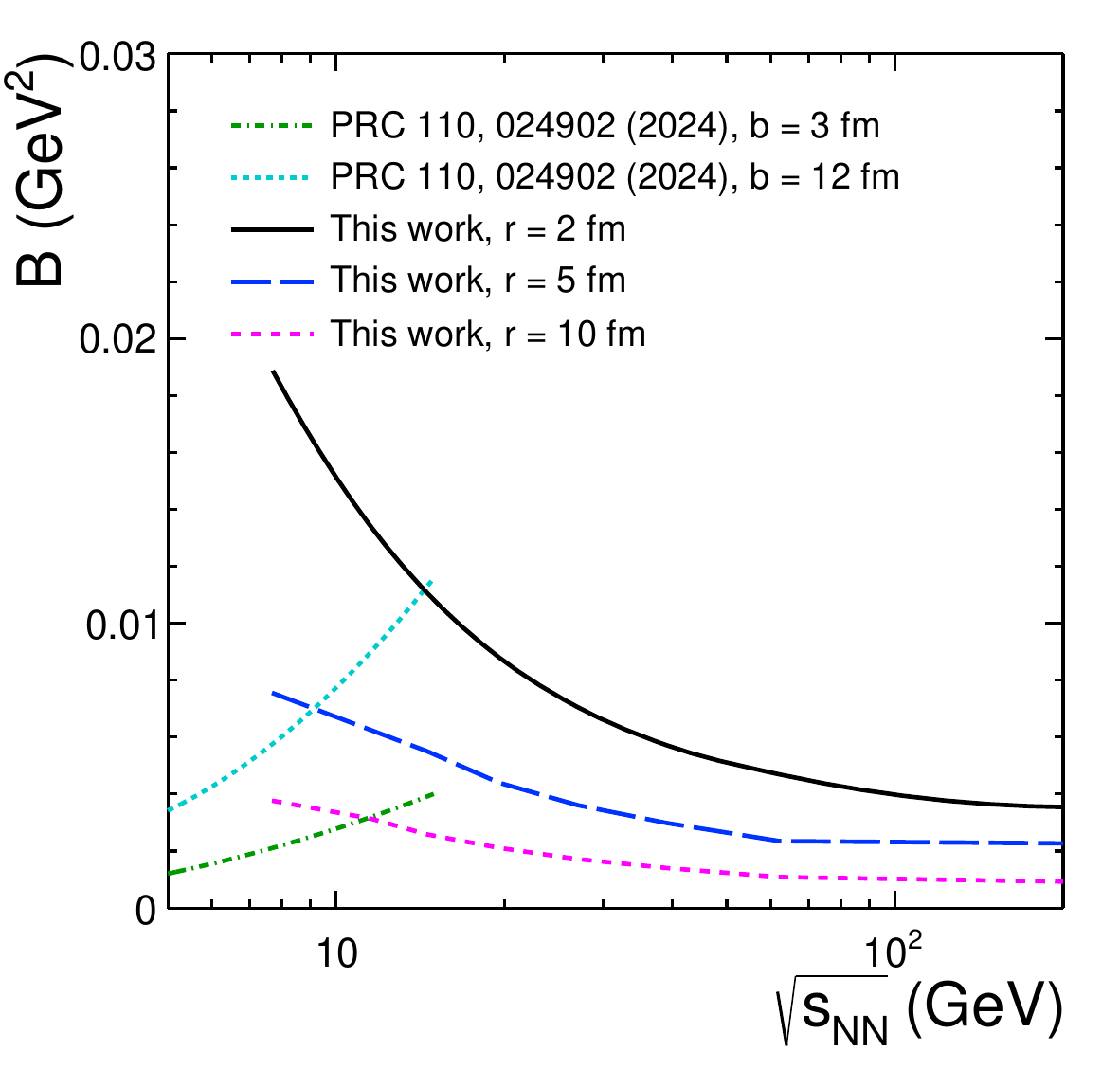}
    \caption{Barnett-induced magnetic field as a function of center of mass energies for the $\omega$ values taken from Ref.~\cite{STAR:2017ckg}, compared with the external magnetic field~\cite{Panda:2024ccj}.}
        \label{fig3}
\end{figure}

For this work, we have taken a smaller hadron sample, which is given in the table~\ref{tab:hadrons}, due to lack of availability of $g$-factor values for all hadronic states. Regardless, the trend will remain almost the same as majority of it is dominated by proton. We have chosen a radius of $r$ = 2 fm for the system, which stems from the fact that for peripheral heavy-ion collisions at LHC and RHIC, the system produced will generally be of the order of a 1 -- 5 fm~\cite{Sahu:2019tch}. However, at all times, the causality relation ($\omega r < 1$) needs to be preserved.

Figure~\ref{fig1} presents the dependence of Barnett magnetization, $M_{\text{Barnett}}$, on the baryochemical potential and temperature for two distinct angular velocity regimes, $\omega = 0.01$ GeV (left panel) and $\omega = 0.05$ GeV (right panel). First, we observe a universal monotonic increase of $M_{\text{Barnett}}$ with both $\mu_B$ and $T$ across both $\omega$ values. Second, the comparison between panels reveals the critical role of rotation strength. The $\omega = 0.05$ GeV case demonstrates significantly enhanced magnetization throughout the entire $(\mu_B, T)$ plane, with particularly pronounced effects in the high-$\mu_B$ region where baryon density maximizes. This $\omega$-dependence confirms the direct proportionality, $M_{\text{Barnett}} \sim \omega$ expected from spin-rotation coupling. The systematic enhancement across all parameter space suggests that Barnett magnetization may dominate over conventional magnetic effects in rapidly rotating QCD matter, especially for neutral hadrons like $\Lambda$ hyperons where Landau quantization is absent.

In Fig.~\ref{fig2} we have plotted the Barnett magnetization of proton and anti-proton exclusively by taking the $\sqrt{s_{\rm NN}}$ dependent $\omega$ values which has been reported in Ref.~\cite{STAR:2017ckg}, which estimates approximate rotation for the system by using the relation, $\omega = k_{B}T(P_{\Lambda} + P_{\bar\Lambda})/\hbar$. Proton gives a positive $M_{\rm Barnett}$ and anti-proton gives a negative $M_{\rm Barnett}$, both the trends decreasing with center of mass energies.

Finally, in fig.~\ref{fig3}, we plot the induced magnetic field for proton which comes from the Barnett effect and compare it with the external magnetic field estimated in Ref.~\cite{Panda:2024ccj} for two different values of impact parameters ($b$). Here, we have chosen three different radius values ($r = 2, 5$ and 10 fm) to observe the system size dependence. The induced magnetic field for protons show a decreasing trend as a function of $\sqrt{s_{\rm NN}}$ for all $r$ values, giving a higher $B_{\rm ind}$ for a smaller $r$. Whereas the external magnetic field show an increasing trend for both the impact parameter, with a higher $B_{\rm ext}$ for higher $b$ value. Here, we have to keep in mind that the $B_{\rm ext}$ is actually the initial magnetic field, with the evolution of the system, it will decline in strength and can be almost negligible when compared to $B_{\rm ind}$, which is the magnetic field estimated at the kinetic freeze-out. This indicates the internally generated $B_{\rm ind}$ surpasses the external $B_{\rm ext}$, suggesting that the Barnett effect may be the dominant source of spin polarization and anomalous transport properties for hadrons with significant magnetic moments in this energy regime.

A critical prediction in the spin physics of heavy-ion collisions is a distinct splitting between the global polarization of $\Lambda$ and $\bar{\Lambda}$ hyperons ($P_{\Lambda} < P_{\bar{\Lambda}}$), which arises from the Barnett effect. While the tremendous fluid vorticity generated in such collisions would, on its own, polarize both particles and antiparticles equally, the Barnett mechanism introduces a crucial dependence on the sign of a particle's magnetic moment. The $\Lambda$ hyperon, with its negative magnetic moment, experiences a suppression of its net polarization, whereas the $\bar{\Lambda}$, with a positive moment, has its polarization enhanced, a signature that aligns with current experimental data. Consequently, the Barnett effect provides a universal mechanism for species-dependent polarization splitting, even among particles with identical spin. Furthermore, since the observed vorticity values are inferred at the kinetic freeze-out after the hadronic phase, the imprint of the Barnett effect persists as a dominant influence, outweighing the impact of the rapidly decaying external magnetic field present in the initial stage of the collision.

\section{Summary}
\label{sec4}

In this work, we present the first study of Barnett magnetization in the hot, dense hadronic matter created in non-central heavy-ion collisions. Using the rotating Hadron Resonance Gas model, we derive the rotation-induced effective magnetic field and compute the resulting magnetization. We find a monotonic increase in magnetization with temperature, baryochemical potential, and angular velocity, with a non-monotonic collision energy dependence arising from the evolving hadronic composition. This magnetization also produces an induced magnetic field which is of the same order as the external magnetic field for protons. This can essentially modify the spin dynamics and transport properties of the QCD matter which has not been previously looked into. Furthermore, this mechanism has profound implications beyond heavy-ion collisions; the extreme rotation of neutron stars could generate colossal Barnett fields, contributing significantly to their magnetization without a conventional dynamo. The competition between the rotation-aligned Barnett field and other external fields could also lead to rich phenomena like spin reorientation transitions, opening new avenues for future research.

\section*{Appendix}

\begin{table}[H]
\centering
\caption{List of hadrons with their magnetic moment and g-factors taken from the PDG list~\cite{ParticleDataGroup:2024cfk}.}
\label{tab:hadrons}
\renewcommand{\arraystretch}{1.5} % Increased row height
\begin{tabular}{|c|c|c|c|c|c|}
\hline
\textbf{Hadron} & \textbf{Mass (GeV)} & \textbf{Charge ($e$)} & \textbf{Spin ($S$)} & \textbf{$\mu/\mu_N$} & \textbf{$g$-factor} \\
\hline
$\pi^0$         & 0.13498  & 0   & 0   & 0       & 0       \\
\hline
$\pi^+$         & 0.13957  & $+1$  & 0   & 0       & 0       \\
\hline
$\pi^-$         & 0.13957  & $-1$  & 0   & 0       & 0       \\
\hline
$K^0$           & 0.49761  & 0   & 0   & 0       & 0       \\
\hline
$\overline{K}^0$ & 0.49761 & 0   & 0   & 0       & 0       \\
\hline
$K^+$           & 0.49368  & $+1$  & 0   & 0       & 0       \\
\hline
$K^-$           & 0.49368  & $-1$  & 0   & 0       & 0       \\
\hline
$\eta$          & 0.54786  & 0   & 0   & 0       & 0       \\
\hline
$\eta'$         & 0.95778  & 0   & 0   & 0       & 0       \\
\hline
$\rho^0$        & 0.77526  & 0   & 1   & 0       & 0       \\
\hline
$\rho^+$        & 0.77526  & $+1$  & 1   & 0       & 0       \\
\hline
$\rho^-$        & 0.77526  & $-1$  & 1   & 0       & 0       \\
\hline
$\omega$        & 0.78265  & 0   & 1   & 0       & 0       \\
\hline
$\phi$          & 1.01946  & 0   & 1   & 0       & 0       \\
\hline
$p$             & 0.93827  & $+1$  & 0.5 & $+2.79$   & $+5.5857$ \\
\hline
$\overline{p}$  & 0.93827  & $-1$  & 0.5 & $-2.79$   & $-5.5857$ \\
\hline
$n$             & 0.93957  & 0   & 0.5 & $-1.91$   & $-3.8261$ \\
\hline
$\overline{n}$  & 0.93957  & 0   & 0.5 & $+1.91$   & $+3.8261$ \\
\hline
$\Lambda$       & 1.11568  & 0   & 0.5 & $-0.61$   & $-1.226$  \\
\hline
$\overline{\Lambda}$ & 1.11568 & 0 & 0.5 & $+0.61$   & $+1.226$  \\
\hline
$\Sigma^+$      & 1.18937  & $+1$  & 0.5 & $+2.46$   & $+4.92$  \\
\hline
$\overline{\Sigma}^-$ & 1.18937 & $-1$ & 0.5 & $-2.46$   & $-4.92$  \\
\hline
$\Sigma^0$      & 1.19264  & 0   & 0.5 & $+0.65$   & $+1.30$   \\
\hline
$\overline{\Sigma}^0$ & 1.19264 & 0 & 0.5 & $-0.65$   & $-1.30$   \\
\hline
$\Sigma^-$      & 1.19745  & $-1$  & 0.5 & $-1.16$   & $-2.32$   \\
\hline
$\overline{\Sigma}^+$ & 1.19745 & $+1$ & 0.5 & $+1.16$   & $+2.32$   \\
\hline
$\Xi^0$         & 1.31486  & 0   & 0.5 & $-1.25$   & $-2.50$   \\
\hline
$\overline{\Xi}^0$ & 1.31486 & 0 & 0.5 & $+1.25$   & $+2.50$   \\
\hline
$\Xi^-$         & 1.32171  & $-1$  & 0.5 & $-0.65$   & $-1.30$   \\
\hline
$\overline{\Xi}^+$ & 1.32171 & $+1$ & 0.5 & $+0.65$   & $+1.30$   \\
\hline
$\Delta^{++}$    & 1.232    & $+2$  & 1.5 & $+4.52$   & $+9.04$  \\
\hline
$\overline{\Delta}^{--}$ & 1.232 & $-2$ & 1.5 & $-4.52$   & $-9.04$  \\
\hline
$\Delta^+$      & 1.232    & $+1$  & 1.5 & $+2.71$   & $+5.42$  \\
\hline
$\overline{\Delta}^-$ & 1.232 & $-1$ & 1.5 & $-2.71$   & $-5.42$  \\
\hline
$\Delta^0$      & 1.232    & 0   & 1.5 & $+0.87$   & $+1.74$  \\
\hline
$\overline{\Delta}^0$ & 1.232 & 0 & 1.5 & $-0.87$   & $-1.74$  \\
\hline
$\Delta^-$      & 1.232    & $-1$  & 1.5 & $-1.90$   & $-3.80$  \\
\hline
$\overline{\Delta}^+$ & 1.232 & $+1$ & 1.5 & $+1.90$   & $+3.80$  \\
\hline
$\Omega^-$      & 1.67245  & $-1$  & 1.5 & $-2.02$   & $-4.04$  \\
\hline
$\overline{\Omega}^+$ & 1.67245 & $+1$ & 1.5 & $+2.02$   & $+4.04$  \\
\hline
\end{tabular}
\end{table}

\section*{Acknowledgment}
 This work has been supported by DGAPA-UNAM PAPIIT No. IG100524 and PAPIME No. PE100124.

\section*{References}

\begin{itemize}
  
%\cite{Barnett:1915uqc}
\bibitem{Barnett:1915uqc}
S.~J.~Barnett,
%``Magnetization by Rotation,''
Phys. Rev. \textbf{6}, 239 (1915).

%\cite{NBE}
\bibitem{NBE}
M.~Arabgol and T.~Sleator, %"Observation of the nuclear Barnett effect,"
Phys. Rev. Lett. 122, 177202 (2019).

%\cite{Kharzeev:2007jp}
\bibitem{Kharzeev:2007jp}
D.~E.~Kharzeev, L.~D.~McLerran and H.~J.~Warringa,
%``The Effects of topological charge change in heavy ion collisions: 'Event by event P and CP violation',''
Nucl. Phys. A \textbf{803}, 227 (2008).

\bibitem{Becattini:2007sr}
F.~Becattini, F.~Piccinini and J.~Rizzo,
%``Angular momentum conservation in heavy ion collisions at very high energy,''
Phys. Rev. C \textbf{77}, 024906 (2008).

%\cite{Endrodi:2013cs}
\bibitem{Endrodi:2013cs}
G.~Endr{\"o}di,
%``QCD equation of state at nonzero magnetic fields in the Hadron Resonance Gas model,''
JHEP \textbf{04}, 023 (2013).

%\cite{Kadam:2019rzo}
\bibitem{Kadam:2019rzo}
G.~Kadam, S.~Pal and A.~Bhattacharyya,
%``Interacting hadron resonance gas model in magnetic field and the fluctuations of conserved charges,''
J. Phys. G \textbf{47}, 125106 (2020).

%\cite{Pradhan:2021vtp}
\bibitem{Pradhan:2021vtp}
G.~S.~Pradhan, D.~Sahu, S.~Deb and R.~Sahoo,
%``Hadron gas in the presence of a magnetic field using non-extensive statistics: a transition from diamagnetic to paramagnetic system,''
J. Phys. G \textbf{50}, 055104 (2023).

%\cite{Sahoo:2023vkw}
\bibitem{Sahoo:2023vkw}
B.~Sahoo, K.~K.~Pradhan, D.~Sahu and R.~Sahoo,
%``Effect of a magnetic field on the thermodynamic properties of a high-temperature hadron resonance gas with van der Waals interactions,''
Phys. Rev. D \textbf{108}, 074028 (2023).

%\cite{Dash:2020vxk}
\bibitem{Dash:2020vxk}
A.~Dash, S.~Samanta, J.~Dey, U.~Gangopadhyaya, S.~Ghosh and V.~Roy,
%``Anisotropic transport properties of a hadron resonance gas in a magnetic field,''
Phys. Rev. D \textbf{102}, 016016 (2020).

%\cite{Dey:2020awu}
\bibitem{Dey:2020awu}
J.~Dey, S.~Samanta, S.~Ghosh and S.~Satapathy,
%``Quantum expression for the electrical conductivity of massless quark matter and of the hadron resonance gas in the presence of a magnetic field,''
Phys. Rev. C \textbf{106}, 044914 (2022).

%\cite{Endrodi:2015oba}
\bibitem{Endrodi:2015oba}
G.~Endrodi,
%``Critical point in the QCD phase diagram for extremely strong background magnetic fields,''
JHEP \textbf{07}, 173 (2015).

%\cite{Endrodi:2024cqn}
\bibitem{Endrodi:2024cqn}
G.~Endrodi,
%``QCD with background electromagnetic fields on the lattice: A review,''
Prog. Part. Nucl. Phys. \textbf{141}, 104153 (2025).

%\cite{Yamamoto:2013zwa}
\bibitem{Yamamoto:2013zwa}
A.~Yamamoto and Y.~Hirono,
%``Lattice QCD in rotating frames,''
Phys. Rev. Lett. \textbf{111}, 081601 (2013).

%\cite{Braguta:2023iyx}
\bibitem{Braguta:2023iyx}
V.~V.~Braguta, M.~N.~Chernodub and A.~A.~Roenko,
%``New mixed inhomogeneous phase in vortical gluon plasma: First-principle results from rotating SU(3) lattice gauge theory,''
Phys. Lett. B \textbf{855}, 138783 (2024).

%\cite{Fukushima:2008xe}
\bibitem{Fukushima:2008xe}
K.~Fukushima, D.~E.~Kharzeev and H.~J.~Warringa,
%``The Chiral Magnetic Effect,''
Phys. Rev. D \textbf{78}, 074033 (2008).

%\cite{Shitade:2020lfe}
\bibitem{Shitade:2020lfe}
A.~Shitade, K.~Mameda and T.~Hayata,
%``Chiral vortical effect in relativistic and nonrelativistic systems,''
Phys. Rev. B \textbf{102}, 205201 (2020).

%\cite{Kharzeev:2010gr}
\bibitem{Kharzeev:2010gr}
D.~E.~Kharzeev and D.~T.~Son,
%``Testing the chiral magnetic and chiral vortical effects in heavy ion collisions,''
Phys. Rev. Lett. \textbf{106}, 062301 (2011).

%\cite{STAR:2023nvo}
\bibitem{STAR:2023nvo}
M.~I.~Abdulhamid \textit{et al.} [STAR],
%``Global polarization of {\ensuremath{\Lambda}} and {\ensuremath{\Lambda}}{\textasciimacron} hyperons in Au+Au collisions at sNN=19.6 and 27 GeV,''
Phys. Rev. C \textbf{108}, 014910 (2023).

%\cite{STAR:2017ckg}
\bibitem{STAR:2017ckg}
L.~Adamczyk \textit{et al.} [STAR],
%``Global $\Lambda$ hyperon polarization in nuclear collisions: evidence for the most vortical fluid,''
Nature \textbf{548}, 62 (2017).

%\cite{ALICE:2022ljz}
\bibitem{ALICE:2022ljz}
S.~Acharya \textit{et al.} [ALICE],
%``Search for the Chiral Magnetic Effect with charge-dependent azimuthal correlations in Xe{\textendash}Xe collisions at sNN=5.44 TeV,''
Phys. Lett. B \textbf{856}, 138862 (2024).

%\cite{Andronic:2005yp}
\bibitem{Andronic:2005yp}
A.~Andronic, P.~Braun-Munzinger and J.~Stachel,
%``Hadron production in central nucleus-nucleus collisions at chemical freeze-out,''
Nucl. Phys. A \textbf{772}, 167 (2006).

%\cite{Karsch:2003vd}
\bibitem{Karsch:2003vd}
F.~Karsch, K.~Redlich and A.~Tawfik,
%``Hadron resonance mass spectrum and lattice QCD thermodynamics,''
Eur. Phys. J. C \textbf{29}, 549 (2003).

%\cite{HotQCD:2014kol}
\bibitem{HotQCD:2014kol}
A.~Bazavov \textit{et al.} [HotQCD],
%``Equation of state in ( 2+1 )-flavor QCD,''
Phys. Rev. D \textbf{90}, 094503 (2014).

%\cite{Fujimoto:2021xix}
\bibitem{Fujimoto:2021xix}
Y.~Fujimoto, K.~Fukushima and Y.~Hidaka,
%``Deconfining Phase Boundary of Rapidly Rotating Hot and Dense Matter and Analysis of Moment of Inertia,''
Phys. Lett. B \textbf{816}, 136184 (2021).

%\cite{Mukherjee:2023ijv}
\bibitem{Mukherjee:2023ijv}
G.~Mukherjee, D.~Dutta and D.~K.~Mishra,
%``Conserved number fluctuations under global rotation in a hadron resonance gas model,''
Eur. Phys. J. C \textbf{84}, 258 (2024).

\bibitem{Pradhan:2023rvf}
K.~K.~Pradhan, B.~Sahoo, D.~Sahu and R.~Sahoo,
%``Thermodynamics of a rotating hadron resonance gas with van der Waals interaction,''
Eur. Phys. J. C \textbf{84}, 936 (2024).

\bibitem{Sahoo:2025fif}
B.~Sahoo, K.~K.~Pradhan, D.~Sahu and R.~Sahoo,
%``Rotational susceptibility of a hot and dense hadronic matter: A possible probe of QCD phase transition,''
[arXiv:2507.03708].

%\cite{Dwibedi:2024amt}
\bibitem{Dwibedi:2024amt}
A.~Dwibedi, N.~Padhan, D.~R.~J.~Marattukalam, A.~Chatterjee, S.~De and S.~Ghosh,
%``Effect of Coriolis Force on Diffusion of D Meson,''
[arXiv:2411.09983].

%\cite{Padhan:2024edf}
\bibitem{Padhan:2024edf}
N.~Padhan, A.~Dwibedi, A.~Chatterjee and S.~Ghosh,
%``Effect of Coriolis force on electrical conductivity tensor for the rotating hadron resonance gas,''
Phys. Rev. C \textbf{110}, 024904 (2024).

%\cite{Andronic:2012ut}
\bibitem{Andronic:2012ut}
A.~Andronic, P.~Braun-Munzinger, J.~Stachel and M.~Winn,
%``Interacting hadron resonance gas meets lattice QCD,''
Phys. Lett. B \textbf{718}, 80 (2012).

\bibitem{griffiths2017electrodynamics}
David J. Griffiths,
\textit{Introduction to Electrodynamics}, 4th edition,
Cambridge University Press, Cambridge, (2017).

%\cite{Cleymans:2005xv}
\bibitem{Cleymans:2005xv}
J.~Cleymans, H.~Oeschler, K.~Redlich and S.~Wheaton,
%``Comparison of chemical freeze-out criteria in heavy-ion collisions,''
Phys. Rev. C \textbf{73}, 034905 (2006).

%\cite{ParticleDataGroup:2024cfk}
\bibitem{ParticleDataGroup:2024cfk}
S.~Navas \textit{et al.} [Particle Data Group],
%``Review of particle physics,''
Phys. Rev. D \textbf{110}, 030001 (2024).

%\cite{Sahu:2019tch}
\bibitem{Sahu:2019tch}
D.~Sahu, S.~Tripathy, G.~S.~Pradhan and R.~Sahoo,
%``Role of event multiplicity on hadronic phase lifetime and QCD phase boundary in ultrarelativistic collisions at energies available at the BNL Relativistic Heavy Ion Collider and CERN Large Hadron Collider,''
Phys. Rev. C \textbf{101}, 014902 (2020).

%\cite{Borsanyi:2013bia}
\bibitem{Borsanyi:2013bia}
S.~Borsanyi, Z.~Fodor, C.~Hoelbling, S.~D.~Katz, S.~Krieg and K.~K.~Szabo,
%``Full result for the QCD equation of state with 2+1 flavors,''
Phys. Lett. B \textbf{730}, 99 (2014).

%\cite{Panda:2024ccj}
\bibitem{Panda:2024ccj}
A.~K.~Panda, P.~Bagchi, H.~Mishra and V.~Roy,
%``Electromagnetic fields in low-energy heavy-ion collisions with baryon stopping,''
Phys. Rev. C \textbf{110}, 024902 (2024).

%\cite{Guo:2019mgh}
\bibitem{Guo:2019mgh}
X.~Guo, J.~Liao and E.~Wang,
%``Spin Hydrodynamic Generation in the Charged Subatomic Swirl,''
Sci. Rep. \textbf{10}, 2196 (2020).

\end{itemize}
\end{document}